\documentclass[manuscript]{aastex}

\usepackage[version=3]{mhchem}
\usepackage{upgreek}
\usepackage{comment}
\usepackage[normalem]{ulem}

\newcommand{\kms}{km\,s$^{-1}$}
\newcommand{\vlsr}{V$_{\rm LSR}$}
\newcommand{\mum}{\ensuremath{\upmu}m}

\slugcomment{Accepted by ApJ May 3, 2016}

\shorttitle{Constraints on SiO gas in HD\,172555 debris disk}
\shortauthors{Wilson, Nilsson, Chen et al.}

\begin{document}


\title{Constraints on the presence of SiO gas in the debris disk of HD\,172555}

\author{T.~L.~Wilson\altaffilmark{1}, R.~Nilsson\altaffilmark{2,3}, C.~H.~Chen\altaffilmark{4}, C.~M.~Lisse\altaffilmark{5}, M.~Moerchen\altaffilmark{4}, H.-U.~K\"{a}ufl\altaffilmark{6}, A.~Banzatti\altaffilmark{4}}

\altaffiltext{1}{United States Naval Research Laboratory, Washington, DC 20375}
\altaffiltext{2}{Astrophysics Department, American Museum of Natural History, Central Park West at 79th Street, New York, NY 10024, USA}
\altaffiltext{3}{Department of Astronomy, Stockholm University, AlbaNova University Center, Roslagstullsbacken 21, 106 91 Stockholm, Sweden}
\altaffiltext{4}{Space Telescope Science Institute, 3700 San Martin Drive, Baltimore, MD 21212}
\altaffiltext{5}{Johns Hopkins Applied Physics Laboratory, Silver Spring MD, 20723}
\altaffiltext{6}{European Southern Observatory, K-Schwarzschild-str. 2, Garching, Germany, 85748}

\begin{abstract}

We have carried out two sets of observations to quantify the properties of SiO gas in the unusual HD\,172555 debris disk: (1) a search for the $J=8-7$ rotational transition from the vibrational ground state, carried out with the APEX 
sub-millimeter telescope and heterodyne receiver at 863\,\mum, and (2) a search at 8.3\,\mum\ for the P(17) ro-vibrational transition of gas phase \ce{SiO}, 
carried out with VLT/VISIR with a resolution, $\lambda/\Delta\lambda$, of 30000.  The APEX measurement resulted in a 3~$\sigma$ non-detection of an interstellar feature, but only an upper limit to emission at the radial velocity and linewidth expected  from HD\,172555.  
The VLT/VISIR result was 
also 
an upper limit.  
These were used to provide limits for the abundance of gas phase \ce{SiO}, for a range of temperatures. The upper limit from 
our 
APEX detection, assuming an 8000\,K primary star photospheric excitation, falls 
 more than an order of magnitude below the self-shielding stability threshold 
derived by \citet{Johnson2012}. 
Our results thus favor a solid-state origin for the 8.3\,\mum\ feature seen in the Spitzer IRS spectrum of the circumstellar excess emission, and the production of 
circumstellar \ion{O}{1} and \ion{Si}{1} by \ce{SiO} UV photolysis. The implications of these estimates are explored in the framework of models of the HD\,172555 circumstellar disk. 

\end{abstract}


\section{Introduction}
The occurrence rate of \emph{debris disks} containing hot dust around sun-like stars is very low, $\sim$2\% 
\citep{Bryden2006a,Bryden2006b,Chen2006}. 
Recent work suggests that the hot dust around most mature
 sun-like stars is transiently regenerated, since it is found to be present in quantities far in excess of 
that 
expected to arise from sublimation of comets \citep{Beichman2005} or slow collisional grinding of 
asteroids left over from the era of planet 
formation \citep{Wyatt2007}. 
%
Terrestrial 
planets are thought to grow by accumulation of smaller objects, through the process of
 collisional accretion \citep{Kenyon2004,Kenyon2006}  
on timescales of 10 to 100\,Myr \citep[see][and references therein]{Wetherill1990,Yin2002,Chambers2004}.

As we will discuss, HD\,172555,  located at a distance of 29.3\,pc and belonging to the $\beta$~Pic Moving Group, is an unusual debris disk system. At an age of more than 15\,Myr 
\citep{Mamajek2014,Bell2015}, it is at a stage of development where the dusty primordial disk should have dissipated and planets formed. 
Yet, HD\,172555 has been identified as a source with pronounced mid-IR excess emission \citep{Schutz2005,Chen2006}, 
which is a signpost of thermally 
radiating circumstellar dust. Dust and gas present at this late stage of disk evolution is likely secondary, i.e.\ produced from bodies 
already aggregated, via 
impacts and cometary outgassing, forming a debris disk. Collisions of planet-sized objects in debris disks may be distinguished from slow, collisional 
grinding since the giant collisions produce much larger masses of warm dust than that expected from steady-state collisional evolution 
\citep{Wyatt2007,Wyatt2011}. 
Unlike most young debris disks, and debris disks in general, 
the  dust surrounding the primary star of HD\,172555 is peculiar. \citet{Lisse2009} took Spitzer/IRS spectra with 
$\lambda/\Delta\lambda = 90$, covering wavelengths $\lambda = 5$--35\,\mum\ and combined  
these results with additional IR photometry 
to determine the Spectral Energy Distribution (SED) from 5\,\mum\ to 100\,\mum. They  
showed that a large amount of fine dust with a temperature of 245\,K was  at 6~AU from the primary, consistent 
with the later measurements of \citet{Pantin2011} using ``Lucky Imaging'', and 
\citet{Smith2012} 
using VLT/MIDI interferometry that showed that the circumstellar dust emission arises from   
 an axisymmetric torus of radius 5.8$\pm$0.6\,AU (0$\farcs$2 at 29.3\,pc). In addition, from blended 8\,\mum\ features \citet{Lisse2009} 
reported that $10^{47}$ \ce{SiO} molecules were present in the gas phase 
and proposed that the most likely 
source of the fine dust and gas phase \ce{SiO} was a massive hypervelocity impact between two rocky planetoids which liberated huge amounts of melted 
and reformed ejecta.  The dust mineralogy is composed primarily of highly refractory, nonequilibrium materials, 
with approximately three
quarters of the Si atoms in silica (SiO$_2$) species. Tektite and obsidian laboratory thermal emission spectra 
(non equilibrium 
glassy silicas found in impact and magmatic systems) are required to fit the data \citep{Lisse2009}. The best-fit 
model size 
distribution for the observed fine dust is a number-size distribution of 
\begin{equation*}
dn/da=a^{-3.95\pm0.10}
\end{equation*}
 While IR photometry of the system has remained 
stable since the 1983 IRAS mission, this steep size distribution, with abundant micron-sized particles, argues for a fresh source of material within the 
last 0.1\,Myr \citep{Lisse2009,Johnson2012}.

In an analysis of the HD\,172555 data, \citet{Johnson2012} argued that the 5.8\,AU torus containing the $\sim 10^{47}$ molecules of gas phase \ce{SiO} 
reported by \citet{Lisse2009} would have to be optically thick in the UV to provide self-shielding stability over decades for any 
circumstellar \ce{SiO} gas, while being optically thin in the IR, to allow detection by Spitzer but reducing any radiation pressure forces on the dust. 
Such a torus should also contain an abundance of silicon and oxygen atoms.  
\citet{Johnson2012} found that about $10^{48}$ atoms of Si and O vapor would be needed to prevent the photodissociation of \ce{SiO}.  HD\,172555 has been re-classified as an A6V--A7V star, which would lower the expected UV flux, and consequently lower the amount needed for stability by about a factor of two.   \ion{O}{1} in 
abundance was detected at 63\,\mum\ by \citet{RiviereMarichalar2012}.  
More recently, \citet{Kiefer2014} reported the detection of time variable \ion{Ca}{2} K and H doublet lines, possibly associated with falling evaporating bodies (FEBs or exocomets), which may provide an alternate explanation for the fine dust and gas phase \ce{SiO} in the HD\,172555 system. 

To refine models for HD\,172555, we have searched for ro-vibrational lines of \ce{SiO} at 8.3\,\mum\ with 
VLT/VISIR with finer velocity resolution than was possible with Spitzer. \footnote{  This publication is partially based on data acquired with observations collected at the European Organization for Astronomical Research in the Southern Hemisphere, Chile (087.C-0669(B)).}   
The VISIR measurements are most sensitive to gas with effective ro-vibrational temperatures up to 500\,K, causing the flux to concentrate in a 
few lines near the ground state. It is also possible that the gas phase \ce{SiO} might be as cold as the dust temperature, $\sim 245$\,K. 
To investigate this possibility, we also searched for the $J=8-7$ rotational transition with the 12-m APEX submillimeter telescope.\footnote{This publication is partially based on data acquired with the Atacama Pathfinder Experiment (APEX). 
APEX is a collaboration between the Max-Planck-Institut fur Radioastronomie, the European Southern Observatory, and the Onsala Space Observatory.} 

\section{Observations and Data Reduction}

\subsection{APEX}

Data were obtained during August, November and December 2014 using the 12-meter Atacama Pathfinder EXperiment (APEX) radio telescope with the APEX-2 heterodyne receiver, which operates in the 267--378\,GHz band.  There are two SiO transitions from the vibrational ground state, at 303.93\,GHz ($J=7-6$) and at 347.33\,GHz ($J=8-7$). We chose the $J=8-7$ transition since this will provide a larger signal for a given column density of SiO gas warmer than 50\,K. There are  thousands of lines from exotic molecular species within $\pm$20\,\kms\ of the center of the spectrometer band (from a perusal of the Splatalogue website)\footnote{See \texttt{ http://www.cv.nrao.edu/php/splat/}}, but all are rather unlikely to be present.  
The $J=8-7$ rotational transition from the vibrational ground state of \ce{SiO} has a rest wavelength of 863\,\mum\ or frequency of $\nu = 347.33$\,GHz. At that frequency, the  full width to half power (FWHP) beam size of the instrument was $17\arcsec$, with a velocity resolution of 1.3\,km\,s$^{-1}$.  The spectrometer was centered on 1\,\kms, the velocity of the star in the Local Standard of Rest (LSR) frame. 
The receiver noise was less than 200\,K and the system noise, including the Earth's atmosphere was 350\,K during the measurements. 
The integration time was 4.8 hours, on-source, with an equal amount of off-source integration time in the  position-switching mode, with an off-source position 10$\arcmin$ from the source position. Calibration was made using the chopper wheel procedure 
\citep[see, e.~g.][]{Wilson2013}. 
The GILDAS software package with CLASS commands was used for reducing and adding data from all scans, performing baseline subtraction, and fitting a Gaussian to the flux at the line position.\footnote{See \texttt{http://www.iram.fr/IRAMFR/GILDAS}.}

\subsection{VLT}

The P(17) transition from the ($v=1$, $J=17$) to the ($v=0$, $J=18$) energy level of \ce{SiO} gas lies in the mid-IR with a rest wavelength of $\lambda = 8.305$\,\mum, or $\nu = 3.61 \times 10^{13}$\,Hz.
Usable VLT/VISIR data were taken on August 19 and 20, 2010, from Paranal in fair weather.

The VISIR spectrometer was used in the high resolution cross-dispersed chopping mode, with a resolution of $\lambda/\Delta\lambda = 3 \times 10^4$, so each channel has a width of 10\,km\,s$^{-1}$. At 8.3\,\mum, the diffraction limited Full Width at Half Power (FWHP) beam size is $0\farcs2$. The total integration time was 3220\,s. The bandpass correction for telluric lines was determined from measurements of $\alpha$\,Cen. The offset from zero level in the VLT data from chopping was set equal to the flux density from the broadband Spitzer spectrum for this wavelength, 1.47\,Jy, to establish the calibration \citep[see Fig.~1 of][]{Johnson2012}. 
The VISIR data were reduced using a suite of IDL routines built to optimize the standard VISIR data
reduction of the ESO Recipe Execution Tool (EsoRex, version 3.9.0), as described in \citet{Banzatti2014}.  
The final  spectrum is shown in Fig.~\ref{fig1}, and the results are discussed further in Section~\ref{sec:results}. 

The line parameters  are listed in Table~\ref{table:results}. Our estimates of the number of \ce{SiO} molecules in the gas phase are derived in 
Section~\ref{sec:results} below. 

\section{Results} \label{sec:results}

\subsection{The APEX Data}
From \citet{Wilson2013}, 
 the relation for a two level system is:
\begin{equation}
\label{general}
N_{\rm l} = 93.5 \frac{g_{\rm l} \, \nu^3}{g_{\rm u} \, A_{\rm ul}} \frac{1}{\left[1-e^{(-4.80 \times 10^{-2} \nu / T_{\rm ex})} \right]}   
\int \tau {\rm d} V,
\end{equation}
where $N_{\rm l}$ is the beam averaged column density in the lower level, $\nu$ is the line frequency in GHz, $g_{\rm l}$ is the 
statistical weight of the lower level, $g_{\rm u}$ of the upper level,  A$_{\rm ul}$ is the Einstein coefficient, $\tau$ is the apparent optical depth and $V$ is the 
velocity width in km\,s$^{-1}$. The excitation temperature,  
 T$_{\rm ex}$, characterizes the populations of the upper and lower levels:
\begin{equation}
\label{pop-2-level}
\frac{n_{\rm u}}{n_{\rm l}} = \frac{g_{\rm u}}{g_{\rm l}} e^{ (-E / T_{\rm ex}) }    , 
\end{equation}
where $n_{\rm u}$ is the population of the upper level, $n_{\rm l}$ the population of the lower level, and $E$ is the 
energy of $n_{\rm u}$ above $n_{\rm l}$ in K. 
For the 863\,\mum\ line, we can simplify  Eq.~\ref{general}  with the use of ${h \nu \ll k T}$: 
%
\begin{equation}
\label{simpler}
N_{\rm l} = 1.95 \times 10^3 \frac{g_{\rm l} \nu^2}{ g_{\rm u} A_{\rm ul}} \int \tau {\rm d} V ,
\end{equation}
 Further, we have 
$T_{\rm MB} = T_{\rm ex} \tau$, $A_{\rm ul}=2.2 \times 10^{-3}$\,s$^{-1}$, $g_{\rm l}$=15,  $g_{\rm u}$=17,  and $\nu=347.333$\,GHz. Then  
 the  column density in the $J=7$ level is given by: 
%
%
 %
\begin{equation}
\label{simpler-1}
N_7=1.0 \cdot 10^{11} \cdot T_{\rm MB} \, \Delta V  
\end{equation}
 The parameters are given in Table 1. A radial velocity of -22 \kms\ is consistent with the interstellar lines of sodium and calcium found by Kiefer et al.~(2014) when observing in the direction of HD 172555. With the detection of a single line, however, no detailed analysis is possible on this potential ISM source.  
 	
In the following, we restrict our analysis to the SiO emission assumed to arise from circumstellar material in the HD172555 system. This should have a \vlsr=1 \kms\ and a FWHP of 33 \kms\ (Gontcharov 2006; Kiefer et al. 2014).  Qualitatively, from a perusal of Fig. 1, no line seems to be present above the noise at this relative velocity. Quantitatively, we employed a Gaussian fit, using \vlsr=1 \kms\ and a FWHP of 33 \kms\, and find an upper limit of T$_{\rm A}^*$=2.3mK, with an RMS spectral noise of T$_{\rm A}^*$=4.63mK. This is an underestimate of the upper limit to any signal, so we have used the 3 RMS noise limit, T$_{\rm A}^*$=13.9mK in the following. From this limit to T$_{\rm A}^*$ and the assumed linewidth, we determined the number of SiO molecules in the 
$J = 7$ rotational level of the vibrational ground state. This is 
\begin{equation}
\label{APEX-value}
N_7 \le 5.5 \times 10^{10}\, {\rm cm}^{-2}
\end{equation}
In the following equations we use equal signs, although this is an upper limit. The total number of \ce{SiO} molecules in all rotational and vibrational levels depends on temperature. This number is estimated using the following analysis. 
The total column density in the ground vibrational state is 
related to the total column density in all rotational levels in the vibrational ground 
state, $N$(total, $v=0$), by: 
\begin{equation}
\label{total-col-den-v-zero}
 N({\rm total}, v=0) = \frac{ N_J}{g_J}  \times {\mathrm e}^{(E_J/T_{\rm rot})} \times  \left( \sum_J g_J \, {\mathrm e}^{E_{\rm rot}/T } \right)  , 
\end{equation}
 where B$_{\mathrm e}$=1.05~K (the molecular constants  of \ce{SiO} are to be found in \citet{Barton2013}. 
Here these are expressed in temperature units. The sum can be replaced by an integral for E$_{\rm rot}$/k greater than 5\,K. \footnote{see 
http://www.exomol.com/exomol2/data/SiO/28Si-16O/EBJT/28Si-16O} Then for $J=7$ we have:
\begin{equation}
\label{APEX-ground-v-state}
{N({\rm total}, v=0)=\frac{N_7}{15} \times \left(  \frac{T}{1.05} \right) \times {\mathrm e}^{(43.6/T)}  },
\end{equation}
 The total column density is the sum over the populations of the vibrational states, given by:
%
\begin{equation}
\label{total-col-den-v-states}
 N({\rm total}, {\rm vibrational})=\sum_{v=0} \, {\mathrm e}^{E_{\rm vib}/T}     
\end{equation}
where E$_{\rm vib}=1758$\,K is the energy of the $v=1$ state above the ground state, in K, and the vibrational levels are very nearly equally 
spaced by this energy. 
%
The column density of \ce{SiO} molecules, in cm$^{-2}$, is the product of Eq.~\ref{APEX-value}, Eq.~\ref{APEX-ground-v-state}, and Eq.~\ref{total-col-den-v-states}. The
total number of \ce{SiO} molecules is the product of the column density and 
and the beam area in cm$^2$. The FWHP beam of APEX at the line frequency is $17\farcs6$; taking this as the diameter, 
the beam area is $4.7 \times 10^{31}$\,cm$^2$.  The upper limit to the total number of \ce{SiO} molecules is plotted as a function of temperature in Fig.~3.  We allow a wide possible range of excitation temperatures, since arguments have been made in the literature by 
 \citet{Lisse2009} for the excitation of \ce{SiO} at 8000\,K, the 
stellar surface temperature, and by \citet{Johnson2012} for an excitation at 245\,K by intermixed dust. 
For two extreme values, 300\,K and 8000\,K, the 3 $\sigma$ limit for  
the total  number of \ce{SiO} molecules is 
$\le$5.5 $\times \, 10^{43}$ and $\le$4.2 $\times \, 10^{45}$, respectively.

\subsection{The VLT Data}
There is no obvious spectral line in the 8.3\,\mum\  VLT/VISIR spectrum, 
so we have used three times the RMS limit in a single spectral channel as the limit.  This is 0.48\,Jy in a 10 km s$^{-1}$ channel. 
For the P(17) transition, the wavelength $\lambda$ is 8.308\,\mum. 
 For the following analysis, we follow the approach used in \citet{Lisse2009}. This can be obtained from Eq.~\ref{general} if $h \nu \ll kT_{\rm ex}$. Using $\mathfrak{J}$, the spontaneous emission coefficient in the equation of radiative transfer, the Einstein $A$ coefficient between the upper level, $J=18$, $v=1$, and lower level, $J=17$, $v=0$:  
\begin{equation}
\label{general-rad-xfer-einstein-A}
\mathfrak{J}_u=\frac{h \nu}{4\pi} \, A_{\rm ul} N
\end{equation}
where $N$ is the column density in units of cm$^{-2}$ and $h$ is Planck's constant. 
\begin{equation}
\label{general-rad-xfer-2}
F = \int S_\nu {\rm d} \nu = \frac{h \nu }{ 4\pi } \, A_{\rm ul} N  \Delta \Omega
\end{equation}
where $N$ is the column density in the upper level (ul). 
The total number of molecules, is estimated from the fraction in the observed  level. For this, the summation (Eq.~\ref{total-col-den-v-zero}) cannot be replaced by an integral. 
%
 For the VLT, the area   subtended by  the FWHP of the diffraction beam  is $6 \times 10^{27}$ cm$^{-2}$
  for a distance of  29.3 pc.  
%
%
The  3 $\sigma$ limit for the flux density in the 8.3 $\mu$ line is $\le$0.48 Jy,   in a 10 \kms\ wide channel. The other parameters are  $v=1$, $J_{\rm u}=17$, $J_{\rm l}=18$, and $A_{\rm ul}=3.44$\,s$^{-1}$, from the \citet{Barton2013} molecular data. 

Estimates of the total number of \ce{SiO} molecules in the gas phase, from our data and from \citet{Lisse2009}, 
are plotted in Fig.~3 as a function of temperature. Our APEX data in Fig.~1 was analyzed using Eq.~\ref{APEX-value}, 
Eq.~\ref{total-col-den-v-zero} and Eq.~\ref{total-col-den-v-states}, under the assumption of LTE.  
%
%
%
%
For $T=8000$\,K, the curve based on the APEX data is more than a factor of more than 40 below the value reported by \citet{Lisse2009}, while the VLT limit is a factor of more than a factor of 10$^3$ below the value of  \citet{Lisse2009}. For $T=300$\,K, the APEX data is more significant, being more than a factor of 10$^3$ below the value of \citet{Lisse2009}.

\section{Discussion}

The 2.3\,\kms wide line at \vlsr=-22 km s$^{-1}$ arises in the 
interstellar medium \citep[see, e.g.,][]{Kiefer2014}, 
so is not related to gas in the disk associated with HD\,172555. Although weak, the radial velocity is in agreement with the value for the interstellar atomic lines found by \citet{Kiefer2014}. 
We will not discuss this feature 
in the following, but note that this result shows that the APEX system has functioned satisfactorally during our observations. In the following, we concentrate on the upper 
limit for \ce{SiO} at 
V$_{\rm LSR}$=0\,km\,s$^{-1}$.  
 
 According to \citet{Johnson2012},  
the abundance 
must exceed 10$^{47}$ to allow survival against photo dissociation of the \ce{SiO} in the 
environment of HD\,172555.  With the re-assessment of the classification of the central star, the amount of \ce{SiO} is lower, but the  VLT data sets  an upper limit much lower than 10$^{46}$ for  T $\ge$ 300~K   
 while the APEX limit 
 is below this value for all temperatures considered. Taken together, these results indicate that the abundance based 
on the feature in the Spitzer spectrum of 
\citet{Lisse2009} should be reduced. 
On the basis of the gas disk 
 analysis of \citet{Johnson2012}, 
it is likely that \ce{SiO} in the gas phase would not survive; instead any \ce{SiO} gas produced by a giant 
hypervelocity impact should be photolyzed into \ion{Si}{1} + \ion{O}{1} atoms by the strong UV flux of the A7V primary, 
producing the \ion{O}{1} population detected by Herschel \citep{RiviereMarichalar2012}. To explain the 8\,\mum\ emission feature seen in the Spitzer data, 
\citet{Johnson2012} suggested an alternative explanation for its source: this would be a shoulder seen from thermal 
emission produced by fine grained, amorphous, hydrated silica (i.e., opalescent) smokes, as seen in laboratory spectra by \citet{Speck2011} 
and as detected on the surface of Mars by \citet{Ruff2011} and \citet{Sun2015}.

If  the abundance of gas phase \ce{SiO} is close to our limit, there are a number of additional findings. 
First, the  dissociated \ce{SiO} gas could be a source of the \ion{O}{1} detected by Herschel, 
 so we do not require substantial \ce{CO} or \ce{CO2} gas to do so. 
The lifetime against photodestruction for \ion{O}{1}  is considerably longer than that of gas phase \ce{SiO}, so the abundance of gas phase \ce{SiO} would have been larger in the past, so the presence of \ion{O}{1}   is consistent with a hypervelocity collision 10$^4$ years ago \citep{Johnson2012}. Alternatively, \citet{RiviereMarichalar2012} speculate that the \ion{O}{1} may have been produced by the slow destruction of grains, that is, collisions, sublimation or sputtering, and remained in place since this species is less affected by radiation pressure than SiO.  
 Second, the heating and vaporization of rocky dust material that produces \ce{SiO} could also produce the \ion{Ca}{2} and \ion{Na}{1} detected by 
\citet{Kiefer2014}, since a process that destroys the \ce{SiO} lattice of silicate dust should also liberate the associated cations in the matrix. 

Our abundance is based on the 3$\sigma$ upper limits from two non-detections. A reasonable follow-up to go deeper and search for a detection of total SiO abundances down to 10$^{42}$, while also mapping the system in the dust continuum, could be provided by a short integration (~ 1 hr, including calibration) with the ALMA array using a resolution of 0.2$''$.  SiO abundances in the ~10$^{42}$ range are expected if the Keifer et al.~(2014) detection of Ca II and Na I are due to the destruction of rocky material in-system. 



\section{Conclusions}
 We have a $3.3\sigma$ detection of the $J=8-7$ rotational line (Fig.~1)  from the interstellar medium toward HD\,172555. For gas associated with HD\,172555 itself, we have   
the $3\sigma$ limits for the $J=8-7$ rotational line and the   P(17) ro-vibration line at 8.3\,\mum\ of gas phase \ce{SiO}, the conclusions are:
\enumerate

\item From the APEX data,  our upper limit for  300\,K, the limit is 5.3 $\times$ 10$^{43}$.  
The APEX data is more sensitive to  cooler gas, so these results preclude large amounts of cooler gas phase \ce{SiO}. For  $T=8000$\,K, the total LTE abundance of gas phase \ce{SiO} is less than 4.4 $\times$ 10$^{45}$. 

\item From the VLT data  our upper limit for the total LTE is less than 2.7 $\times$ 10$^{47}$ at 8000\,K.   At lower temperatures, the abundance is smaller. See Fig.~3 for the dependence of abundance on temperature. 

\item   The upper limit obtained with  APEX, if interpreted as representative of the minimum past  \ce{SiO} gas content of the system, does not exclude \ce{SiO} as a source of the  \ion{O}{1} detected with Herschel, 
so substantial amounts of \ce{CO} or \ce{CO2} gas are not required. 


\item  The heating and vaporization of 
rocky dust material that could produce gas phase \ce{SiO} to the upper limit found with APEX 
could also produce the \ion{Ca}{2} and \ion{Na}{1} detected by \citet{Kiefer2014}.

\item  With the \ce{SiO} gas stability analysis of Johnson et al.~(2012), revised down to about 5 $\times$ 10$^{48}$ molecules for the approximately two times lower UV flux for an A6V type primary star, our upper limit on \ce{SiO} gas suggests that the feature in the Spitzer spectrum previously attributed to gas phase \ce{SiO} is caused by an amphorous silica compound.  





\acknowledgments{We thank an anonymous referee for helpful comments on the text and interpretation. R.N. was funded by the Swedish Research Council's International Postdoctoral Grant No.~637-2013-474.}


\facility{APEX (APEX-2), VLT (VISIR)}.



\clearpage

\begin{figure}
\epsscale{.80}
\plotone{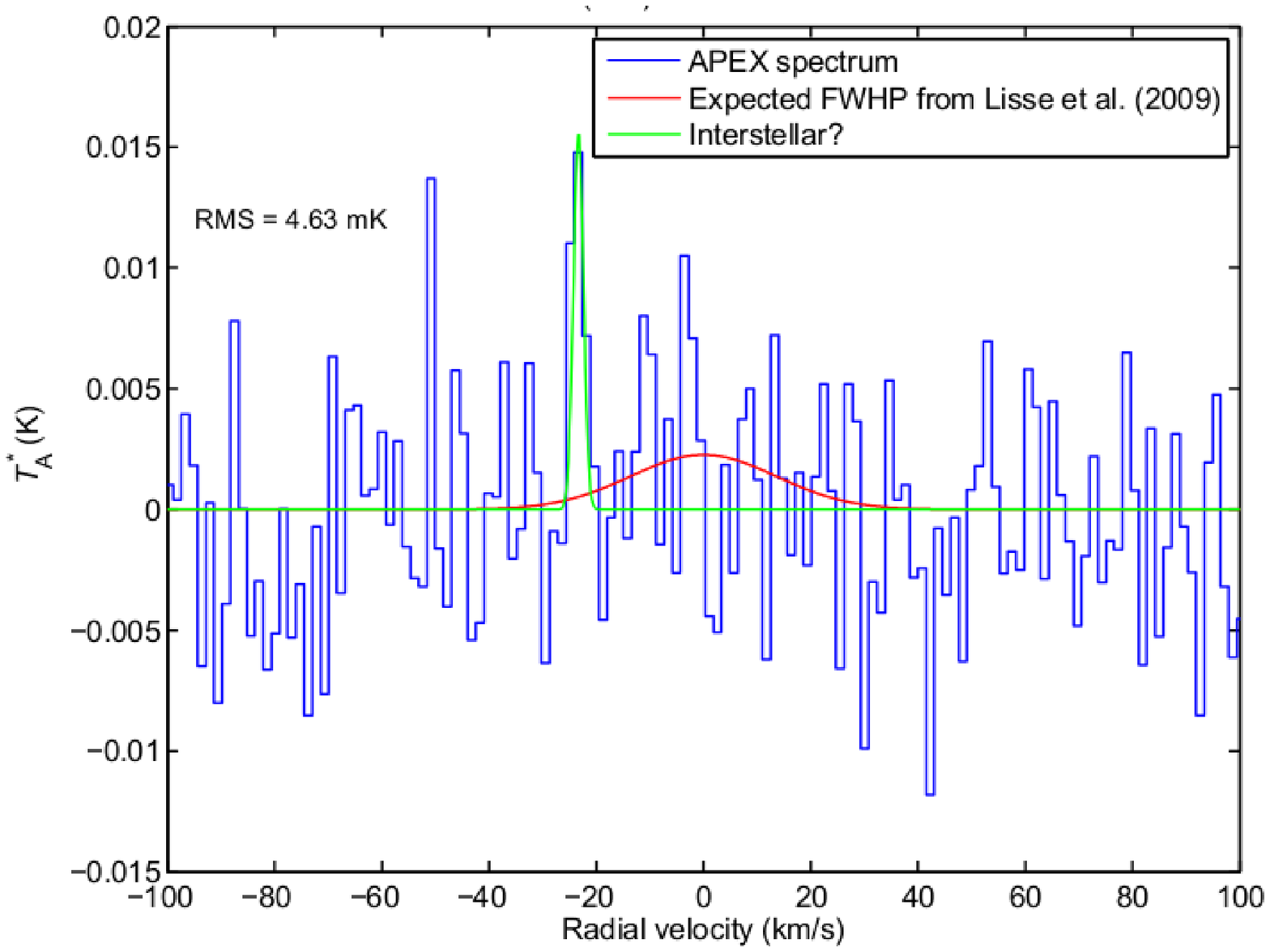}
\caption{  The spectrum taken with the APEX millimeter/sub-mm telescope. The fit in green shows the 3.3~$\sigma$ feature which is taken as a  detection of the  SiO $J=8-7$ rotational transition from the interstellar medium, at V$_{\rm LSR}$=-22~km s$^{-1}$. The the fit in red was constrained to be centered at \vlsr=1~km s$^{-1}$ and  to have a width of 33~km s$^{-1}$ This spectrum is on a T$_{\rm A}^*$ scale. The peak T$_{\rm A}^* \le$ 2.3 mill~K. To 
convert to  main-beam brightness temperatures, multiply by 1.3. For the estimate of the total number of SiO molecules, a T$_{\rm A}^*$ equal to three times RMS, 13.9 milli~K, was used.  The radial velocities are with respect to the Local Standard of Rest using the Standard Solar Motion. }
\label{fig2}
\end{figure}

\begin{figure}
\epsscale{.80}
\plotone{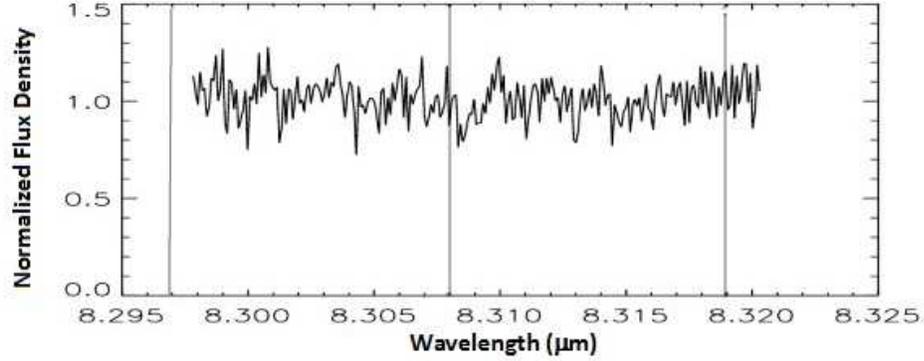}
\caption{The spectrum of the P(17) ro-vibrational line of gas phase \ce{SiO}, taken with VLT/VISIR with a resolution of 3$\times$10$^4$. 
The central vertical line at 8.308\,\mum\ shows the expected location of the P(17) line of SiO. }
\label{fig1}
\end{figure}

\begin{figure}
\epsscale{.80}
\plotone{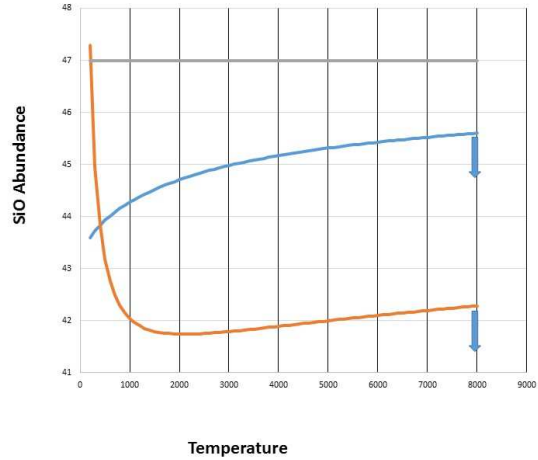}
\caption{A plot of  limits for 
the total abundance of gas phase SiO in HD172555 as a function of the LTE temperature.  For T$\ge$500~K, the lowest curve is based on our VLT measurement of the ro-vibrational P(17) line at 8.3\,\mum.  The next higher curve is based on our APEX measurement of the J=8-7 rotational transition at 347.33 GHz. 
 The downward arrows indicated that these values are upper limits.  The horizontal line shows 
the abundance of gas phase SiO reported by Lisse et al.~2009, from their Spitzer data. 
}
\label{fig3}
\end{figure}

 \begin{deluxetable}{ccccccc}
 \tablecaption{Results}
 \tablehead{ \colhead{Telescope} & \colhead{Transition} & \colhead{Wavelength } & \colhead{$S_{\nu}$} & \colhead{$T_{\rm MB}$} & \colhead{$\Delta {\rm V_{1/2}}$}  &  \colhead{$\Delta {\rm V_{LSR}}$}  \\
\colhead{} &\colhead{} &\colhead{(\mum)}  &   \colhead{(Jy)}          &   \colhead{(K)}        &      \colhead{(km s$^{-1}$)} &   \colhead{(km s$^{-1}$)} \\
 \colhead{(1)}  & \colhead{(2)}  &        \colhead{(3)}      &      \colhead{(4)}           &  \colhead{(5)}   &  \colhead{(6)} &  \colhead{(7)}  }
 \startdata
VLT  &  P(17) &8.3             &   $\le$0.48$^a$      &      \nodata                 & 10$^b$  &  0$^c$\\
APEX & $ J=8-7$ &863           &     \nodata          &  0.018$\pm$0.006~K           & 2.63$^d$   & -22$^e$\\
APEX & $J=8-7$  &863           &     \nodata          & $\le$0.017$^a$               & 33$^f$     &  0$^c$ \\
 \enddata
 \tablenotetext{a}{\, 3$\sigma$ limit}
 \tablenotetext{b}{\, Velocity resolution}
 \tablenotetext{c}{\, From gas associated with HD~172555}
 \tablenotetext{d}{\, Measured line width}
 \tablenotetext{e}{\, From interstellar gas}
 \tablenotetext{f}{\, Assumed line width}
 \label{table:results}
 \end{deluxetable}

\end{document}